\documentclass[doublecol]{epl2} 

\title{Reply to the Comment by S. Su \etal}

\author{Y. Apertet\inst{1} \and H. Ouerdane\inst{2}\inst{3} \and O. Glavatskaya\inst{2}\inst{4} \and C. Goupil\inst{2}\inst{3} \and P. Lecoeur\inst{1}}
\shortauthor{Y. Apertet \etal}

\institute{                    
  \inst{1} Institut d'Electronique Fondamentale, Universit\'e Paris-Sud, CNRS, UMR 8622, F-91405 Orsay, France\\
  \inst{2} Laboratoire CRISMAT, UMR 6508 CNRS, ENSICAEN et Universit\'e de Caen Basse Normandie, 6 Boulevard Mar\'echal Juin, F-14050 Caen, France\\
  \inst{3} Universit\'e Paris Diderot, Sorbonne Paris Cit\'e, Institut des Energies de Demain (IED) URD 0001, 75205 Paris, France\\
  \inst{4} Renault SAS, SAS FR TCR AVA 058, 1 avenue du Golf, 78288 Guyancourt, France
}
\pacs{84.60.Rb}{Thermoelectric energy conversion}
\pacs{85.80.Fi}{Thermoelectric devices}
\pacs{88.05.De}{Thermodynamic constraints}

\abstract{}

\begin{document}

\maketitle

In our Letter ~\cite{Apertet1}, we showed how optimal working conditions for thermoelectric generators (TEG) with realistic thermal coupling to heat baths may be achieved. In their comment \cite{CommentSu}, Su and co-workers claim that our results and conclusions \cite{Apertet1} are flawed because:\\

\noindent $i/$ the maximum output power of the TEG $P_{\rm max}$ is a monotonically increasing function of the ratio of the heat exchangers' thermal conductances to the open-circuit thermal conductance of the TEG: $K_{\rm contact}/K_{_{I=0}}$, which is in contradiction with Fig.~2 of Ref.~\cite{Apertet1}\\

\noindent $ii/$ $K_{_{I=0}}$ could not be used as a variable.\\

\noindent $iii/$ Eq.~(21) in Ref.~\cite{Apertet1} is erroneous.\\

We believe that the first two observations stem from a misunderstanding of the condition for which power and efficiency optimizations are performed. The third remark raises a point that we must clarify here. The notations used in the following discussion are the same as those adopted in Ref.~\cite{Apertet1}.

\section{On the choice of the variables}
In Ref.~\cite{Apertet1}, we study the impact of non-ideal heat exchangers on the TEG performances. We show in particular that the thermal matching between the heat exchangers' thermal conductances, $K_{\rm contact}$, and that of the TEG, $K_{\rm TEG}$, yields power maximization. We highlight however that this condition only applies to cases where $K_{\rm contact}$ is fixed: the variable used to perform the optimization is $K_{_{I=0}}$ only. Therefore, though the maximum power $P_{\rm max}$ in Fig.~(2) is plotted against the normalized dimensionless ratio $K_{_{I=0}}/K_{\rm contact}$, this ratio cannot be considered as the optimization variable: only $K_{_{I=0}}$ is. We thus rebut point $i/$ of Su and co-worker's criticism.

We analyze power maximization for TEGs with different figures of merit $Z\overline{T}$ using $K_{_{I=0}}$ as a variable. For a fixed value of $Z\overline{T}$, we adapt the value of the Seebeck coefficient $\alpha$ for each value of $K_{_{I=0}}$: $\alpha  = \sqrt{ Z R K }$. This method is adapted from the article from Nemir and Beck \cite{nemirbeck}, in which the authors discuss the significance of the figure of merit. Contrary to the statement of Su and co-workers in their comment \cite{CommentSu}, one should not consider that $\alpha^2 / R$ is a fixed parameter; so using $K_{_{I=0}}$ as a variable is not a mistake. The variation of $\alpha$ and $K_{_{I=0}}$ for a constant $Z\overline{T}$ allows to put forth the fact that the \emph{individual} value of the different parameters may be as important for a TEG's performance as the value of the \emph{global} figure of merit $Z\overline{T}$ when realistic thermal contacts are accounted for. We thus prove wrong point $ii/$ of Su and co-worker's criticism.

\section{On the definition and optimisation of the efficiency} 
In Ref.~\cite{Apertet1}, we used approximations in order to make the derivations tractable. The main hypothesis is that the electrical power delivered by the TEG is very small compared to the average thermal current $I_{Q}$, i.e., the efficiency is low. This allows to use a thermal anologue of the potential divider formula to determine the temperature difference experienced by the TEG. We then get $I_{Q} \approx I_{Q_{\rm in}} \approx I_{Q_{\rm out}}$. Consequently, in order to keep the expression of the efficiency $\eta$ straightforward we defined it as $\eta = P/I_{Q}$ instead of $\eta = P/I_{Q_{\rm in}}$. Although this derivation is coherent with the hypothesis made before, a statement that this expression is only a reasonable approximation is missing. We thank the Authors of the Comment to point out this omission.

In order to evaluate the difference between this approximation and the correct definition for the efficiency, we rewrite Eq.~(8) of Ref.~\cite{CommentSu}:

\begin{equation}
\frac{P}{I_{Q}} = \frac{P}{ \left( \dot{Q}_{\rm in} + \dot{Q}_{\rm out} \right) / 2} = \frac{\eta}{ 1 - \eta / 2}
\end{equation}

\noindent The correction factor is then $1/(1 - \eta/2)$, which reduces to $1$ when $\eta$ is small compared to $1$.

\begin{figure}
\scalebox{.34}{\includegraphics*{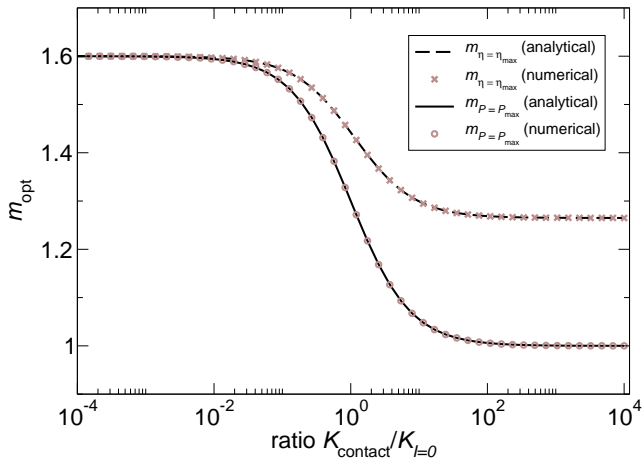}}
\caption{\label{fig1}Variations of the optimal parameters $m_{_{\eta=\eta_{\rm max}}}$ and $m_{_{P=P_{\rm max}}}$ as functions of $K_{\rm contact}$ scaled to $K_{_{I=0}}$. Comparison between the analytical results derived in Ref.~\cite{Apertet1} and exact numerical calculations.}
\end{figure}

Finally to demonstrate that the use of this approximation does not change the results and conclusion of Ref.~\cite{Apertet1}, we check that the analytical expressions for $m_{_{\eta=\eta_{\rm max}}}$ and $m_{_{P=P_{\rm max}}}$ given by Eqs.~(14) and (19) of Ref.~\cite{Apertet1} are in excellent agreement with the exact numerical results obtained using Eq.~(9) of Ref.~\cite{Apertet1}. The two parameters are ahown as functions of $K_{\rm contact}$ scaled to $K_{_{I=0}}$ in Fig.~\ref{fig1}. The values of the thermoelectric parameters for this example are taken from Ref.~\cite{nemirbeck}. We see that the analytical expressions match perfectly the values for both $m_{_{\eta=\eta_{\rm max}}}$ and $m_{_{P=P_{\rm max}}}$ obtained numerically: all the results and conclusions in Ref.~\cite{Apertet1} remain valid.

\section{Summary}
We showed that the criticisms made by Su and co-workers concerning the choice of the variables are unsubstantiated, and we provided additional justification for the use of the approximation made in Eq.~(21) of Ref.~\cite{Apertet1} to derive the electrical condition leading to performance optimization.

\end{document}